\documentstyle[aps,multicol]{revtex}
\include{epsf}

\begin{document}
\title{Elastic scattering loss of atoms from colliding Bose-Einstein
condensate wavepackets}

\author{Y.\ B.\ Band$^{\,1,2}$ Marek Trippenbach$^{\,2,3}$, 
J.\ P.\ Burke Jr.$^{\,2}$, and P.\ S.\ Julienne$^{\,2}$}

\address{${}^{1}$ Department of Chemistry,
Ben-Gurion University of the Negev, Beer-Sheva, Israel 84105 \\
${}^{2}$ Atomic Physics Division, A267 Physics,
National Institute of Standards and Technology, Gaithersburg, MD  
20899\\
${}^{3}$ Institute of Experimental Physics, Optics Division, 
Warsaw University, ul.~Hoza 69, Warsaw 00-681, Poland}
\maketitle

\begin{abstract}
Bragg diffraction of atoms by light waves has been used to create high
momentum components in a Bose-Einstein condensate.  Collisions between
atoms from two distinct momentum wavepackets cause elastic scattering
that can remove a significant fraction of atoms from the wavepackets
and cause the formation of a spherical shell of scattered atoms.  We
develop a slowly varying envelope technique that includes the effects
of this loss on the condensate dynamics described by the
Gross-Pitaevski equation.  Three-dimensional numerical calculations
are presented for two experimental situations: passage of a moving
daughter condensate through a non-moving parent condensate, and
four-wave mixing of matter waves.
\end{abstract}


\begin{multicols}{2}
\narrowtext

\begin{figure}
\centerline{\epsfxsize=3.0in\epsfbox{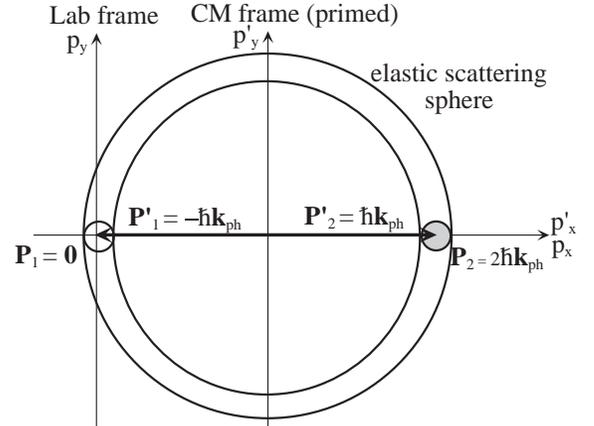}}
\caption {(a) Wavepackets in momentum space with laboratory frame
central momenta ${\bf P}_1 = {\bf 0}$ and ${\bf P}_2 = 2\hbar{\bf
k}_{ph}$ undergoing elastic scattering to produce a spherical shell of
elastically scattered particles.  In the CM frame moving with velocity
${\bf v}_{cm} = \hbar{\bf k}_{ph}/m$, momenta are shifted by
$-\hbar{\bf k}_{ph}$.}
\label{f1}
\end{figure}

A light-induced potential applied to a Bose-Einstein condensate (BEC)
can be used to make high momentum daughter BEC wavepackets which
propagate through the parent condensate
\cite{Ovchinnikov99,Kozuma99,Stenger}.  High momentum means very large
in relation to the mean momentum in the parent wavepacket and the
momentum $mv_s$ where $v_s$ is the sound velocity in the parent.  Such
techniques have been used to make an atom laser \cite{nist_al}, to
study the coherence properties of condensates
\cite{Stenger,nist_cl_exp,cl_theory}, and to study nonlinear four-wave
mixing (4WM) of coherent matter waves \cite{4WM_1,4WM_2}.  As
explained in this Letter, elastic scattering between condensate atoms
from different momentum wavepackets can remove copious numbers of
atoms from these moving wavepackets.  Recently, profuse elastic
scattering of atoms between daughter and parent BEC wavepackets has
been observed at MIT \cite{SK-K}.  Such losses will be an important
consideration for atom optics applications.  Fig.~\ref{f1}
schematically shows wavepackets in momentum space where the high
momentum wavepacket with central momentum ${\bf P}_2 = 2 \hbar {\bf
k}_{ph}$ was produced by optically-induced Bragg scattering from the
${\bf P}_1 =0$ initial wavepacket.  Here $\hbar k_{ph}=h/\lambda_{ph}$
is the photon momentum for light with wavelength $\lambda_{ph}$.  The
spherical shell in Fig.~\ref{f1} (excluding the condensate
wavepackets) results from elastic scattering between atoms from the
${\bf P}_1$ wavepacket and atoms from the ${\bf P}_2$ wavepacket.  The
elastically scattered atoms in the spherical shell can neither be
described as part of the mean-field of the BEC, nor can the formation
\cite{mom_considerations} or evolution of the spherical shell be
modeled using the usual Gross-Pitaevskii equation (GPE)
\cite{Dalfovo}, Eq.~(\ref{GP}).  In what follows we use the term
``elastic scattering'' to mean {\it only} those non-forward elastic
scattering processes not accounted for within the GPE.

Here we provide a simple means of describing the loss of atoms from
the condensate wavepackets due to the elastic scattering mechanism. 
This is made possible by (a) using the appropriate momentum dependence
of the nonlinear coupling constant in the GPE
\cite{Huang,Stoof-Prokakis} and (b) using a newly developed
slowly-varying-envelope approximation (SVEA) for the condensate
wavefunction in systems with both slow and fast momentum components
\cite{Trip4WM}.  Since the SVEA treats each distinct momentum
wavepacket separately, we can incorporate the correct momentum
dependence in the nonlinear coupling constants.  Thus, we can treat
the effect of elastic scattering losses on the condensate dynamics
using the SVEA version of the GPE, even in single spin component
systems.  We first outline the theory for describing elastic
scattering loss and then present two examples, one applied to output
coupling of atom laser pulses from a BEC source, and the other to 4WM.
A more complete discussion of the theory and further applications will
be presented elsewhere \cite{TheoryPaper}.
 
The GPE for a single spin component BEC at zero temperature can be
written as \cite{Dalfovo}
\begin{equation}
i\hbar \frac{\partial\Psi}{\partial t} = (T_{{\bf x}}+ V({\bf x},t) +
U_0 N_{T} |\Psi|^2) \Psi \ , \label{GP}
\end{equation}
where $T_{{\bf x}} = \frac{-\hbar^2}{2m} \nabla^2$ is the kinetic
energy operator, $V({\bf x},t)$ is the external trapping potential
imposed on the atoms, $N_{T}$ is the total number of atoms in the
condensate, and $\Psi$ is normalized to unity.  Although the coupling
constant is usually expressed as $U_{0} = \frac{4\pi \hbar^{2}
a_{0}}{m}$, where $a_0$ is the $s$-wave scattering length and $m$ is
the atomic mass, it is more correct at zero temperature to express
$U_{0}$ in terms of the many-body $T$-matrix, which is often
well-approximated by the 2-body $T$-matrix
\cite{Huang,Stoof-Prokakis}: $U_{0}(k) = -\frac{4\pi \hbar^{2}
T(k)}{mk}$.  Here $T=\frac{S-1}{2i}$, $S$ is the unitary $S$-matrix,
and $\hbar k$ is the relative momentum of the colliding atoms.  If we
assume no inelastic scattering and expand $T(k)/k$ in powers of $k$
using $S(k)=e^{-2ika_0}+O(k^3)$, we obtain:
\begin{equation}
  \frac{T(k)}{k} = -a_0 +i k a_0^2 + O(k^2) \,.
  \label{Tmatrix}
\end{equation}
In a normal condensate, the lead term in Eq.~(\ref{Tmatrix}) gives the
usual GPE. However, if two wavepackets with very different momenta
interact (cross-energy terms in Eqs.~(\ref{SVEA1}-\ref{SVEA2})),
$ka_{0}$ need not be negligible ($ka_{0}=0.06$ for the Na condensate
example given below), and the second term in Eq.~(\ref{Tmatrix}) must
be taken into account.  This term generates the elastic scattering
loss term in the GPE. If we apply the optical theorem to the forward
scattering amplitude, it is this second term that is responsible for
scattering out of the forward direction.  Suitable generalizations of
Eq.~(\ref{Tmatrix}) allow inclusion of inelastic collision losses and
treatment of multiple spin-component condensates.

The application of an optical standing wave pulse diffracts a fraction
of the initial condensate into a high momentum component
\cite{Ovchinnikov99,Kozuma99,Stenger}.  To a very good approximation,
the wavefunction immediately after application of a set of optical 
pulses is given by the superposition of wavepackets,
\begin{equation}
\Psi({\bf x},t=0) = \psi({\bf x})  \sum_{j=1}^{J}
a_{j} \exp(i{\bf k}_{j}\cdot{\bf x}) \ ,   \label{in_con}
\end{equation}
where $J$ is the number of distinct momentum wavepackets present. 
Here $\psi({\bf x})$ is the initial wavefunction of the parent
condensate before application of the optical pulses; it is the
solution to the GPE with a harmonic potential centered around ${\bf
x}={\bf 0}$.  We assume the momentum differences $\hbar|{\bf k}_{i} -
{\bf k}_{j}|$ to be much larger than both the momentum spread in the
initial parent BEC wavepacket and the momentum $mv_s$ associated with
the speed of sound $v_s$ in the BEC (hence, superfluid suppression of
collisions \cite{SK-K} does not occur).  Since different wavepackets
do not overlap in momentum space, $\sum_{j=1}^J |a_{j}|^{2} = 1$.

The SVEA is made by writing the wavefunction as
\begin{equation}
\Psi({\bf x},t) = \sum_{j} \Phi_j({\bf x},t) \exp(i{\bf k}_j{\bf x} -
i\omega_j t) \,.
\label{SVEAPsi}
\end{equation}
Eq.~(\ref{SVEAPsi}) explicitly separates out the fast oscillating
phase factors representing central momentum $\hbar{\bf k}_j$ and
kinetic energy $E_j = \hbar\omega_j = \hbar k_j^2/2m$ and defines the
slowly varying envelopes $\Phi_j$, which vary in time and space on
much slower scales than the phases.  Consequently, full
three-dimensional (3D) calculations of the envelopes is numerically
tractable, as we describe in more detail elsewhere
\cite{cl_theory,Trip4WM}.  In the first example we consider below, we
take only two components, i.e., $j=1,2$ and the initial condition is
$\Phi_j({\bf x},t=0)=a_{j} \psi({\bf x})$.  In the 4WM process,
$j=1,\ldots,4$, with $\Phi_j({\bf x},t=0) = a_{j} \psi ({\bf x})$ for
$j=1,2,3$, and the $j = 4$ envelope is initially unpopulated,
$\Phi_4({\bf x},t=0)=0$.  This envelope becomes populated as a result
of the coherent 4WM process.  Substituting the SVEA form for the
wavefunction into the GPE, collecting terms with the same phase
factors, multiplying by the complex conjugate of the appropriate phase
factors, and neglecting terms that are not phase matched (those for
which momentum and energy are not conserved) we obtain a set of
coupled SVEA equations for $\Phi_j({\bf x},t)$:
\begin{eqnarray}
&&\left( \frac{\partial}{\partial t} + (\hbar {\bf k}_j/m) \cdot {\bf
\nabla} + \frac{i}{\hbar}(-\frac{\hbar^{2}}{2m}\nabla^{2} + V({\bf
r},t) ) \right) \Phi_j = \nonumber \\
&&-\frac{i}{\hbar} U_0 N_{T} \sum_{qrs} \delta ({\bf k}_{jqrs})
\delta(\omega_{jqrs}) \Phi_q \Phi_r^{*} \Phi_s \ .
\label{SVEA}
\end{eqnarray}
Only phase-matched terms, for which ${\bf k}_{jqrs} = {\bf k}_j - {\bf
k}_q + {\bf k}_r - {\bf k}_s =0$ and $\omega_{jqrs} = \omega_j -
\omega_q + \omega_r - \omega_s =0$, are retained on the right hand
side of Eqs.~(\ref{SVEA}).

For simplicity, we consider explicitly the SVEA equations for the case
where only two central momentum components, ${\bf 0}$ and $2\hbar{\bf
k}_{ph}$, are present.  Then, only ``phase-modulation'' nonlinear
self- and cross-energy interaction terms are present, as opposed to
the case when three central momentum components are present and 4WM
terms also arise.  It is convenient to go to a center of mass frame
moving with velocity ${\bf v}_{cm}=\hbar{\bf k}_{ph}/m$ (see
Fig.~\ref{f1}).  In this frame, wavepacket $1$ has momentum
$-\hbar{\bf k}_{ph}$, wavepacket $2$ has momentum $\hbar{\bf k}_{ph}$,
and elastic scattering from the $\pm \hbar{\bf k}_{ph}$ wavepackets
creates a spherical shell expanding in $4\pi$ steradians with momentum
$|\hbar {\bf k}_{elas}| = |\hbar {\bf k}_{ph}|$.  The SVEA equations
in this frame are given explicitly by:
\begin{eqnarray}
&& \left( \frac{\partial}{\partial t} + (-{\bf v}_{cm}) \cdot
{\bf \nabla} + \frac{i}{\hbar}\left(-\frac{\hbar^{2}}{2m}\nabla^{2} +
V({\bf r},t)\right ) \right) \Phi_1 = \nonumber \\
&& -i\frac{4\pi \hbar a_0}{m} N_{T}
(|\Phi_1|^2 + 2|\Phi_2|^2) \Phi_1 
- \frac{(v_{rel}) \sigma N_T}{2} |\Phi_2|^2 \Phi_1 \ ,
\label{SVEA1}
\end{eqnarray}
\begin{eqnarray}
&& \left( \frac{\partial}{\partial t} + {\bf v}_{cm} \cdot
{\bf \nabla} + \frac{i}{\hbar}\left(\frac{-\hbar^{2}}{2m}\nabla^{2} +
V({\bf r},t)\right ) \right) \Phi_2 = \nonumber \\
&& -i\frac{4\pi \hbar a_0}{m} N_{T} 
(|\Phi_2|^2 + 2|\Phi_1|^2) \Phi_2 
- \frac{(v_{rel}) \sigma N_T}{2} |\Phi_1|^2  \Phi_2 \, .
\label{SVEA2}
\end{eqnarray}
where $\sigma = 8\pi a_0^2$ and $v_{rel}=2\hbar k_{ph}/m = 2v_{cm}$ is
the relative velocity of the two wavepackets.  The factor of $2$
multiplying the nonlinear cross-energy interaction terms results from
expanding $|\Psi|^2 \Psi$ with $\Psi({\bf x},t) = \sum_{j=1}^2
\Phi_j({\bf x},t) \exp(i{\bf k}_j{\bf x} - i\omega_j t)$ and
collecting the phase-matched terms appropriately.  In the self-energy
term proportional to $|\Phi_i|^2 \Phi_i$, $i=1,2$, only the lead term
in Eq.~(\ref{Tmatrix}) is retained.  However, both terms in
Eq.~(\ref{Tmatrix}) are retained in the cross-energy terms $|\Phi_i|^2
\Phi_j$, $i=1,2$, $j=2,1$, leading to the elastic scattering loss
terms proportional to $v_{rel}\sigma$.

The form of the elastic collisional loss terms can also be motivated 
by a classical hydrodynamic picture of elastic collisions
of a cloud of atoms having a central velocity ${\bf v_1} = {\bf
P_1}/m$ and density $n_1({\bf x},t)$, with a cloud of atoms having
velocity ${\bf v}_2$ and density $n_2({\bf x},t)$.  The atomic
densities can be determined from:
\begin{equation}
\frac{\partial n_1({\bf x},t)}{\partial t} + {\bf v_1} \cdot {\bf
\nabla} n_1({\bf x},t) = - |{\bf v_1}-{\bf v}_2| \sigma n_1 n_2 \ , 
\label{dn/dt}
\end{equation}
and a similar equation for $\frac{\partial n_2({\bf x},t)}{\partial
t}$.  Thus, elastic scattering takes atoms out of both clouds.  For
sufficiently slow relative atomic velocities so that only $s$-wave
scattering occurs, and both atoms have the same spin quantum numbers,
$\sigma = 8\pi a_0^2$.  Here the relative velocity is $v_{rel}=|{\bf
v_1}-{\bf v}_2| = 2v_{cm}$.  The extra factor of $\frac{1}{2}$ in the
loss terms in Eqs.~(\ref{SVEA1})-(\ref{SVEA2}) are due to the fact
that these are equations for amplitudes ($\Phi$), not densities
($|\Phi|^2$).  Inelastic scattering with cross section $\sigma_{in}$,
if present, is easy to include by replacing $\sigma$ by
$\sigma+\sigma_{in}$.

If we apply this theory to condensates with two spin components, the
cross section which appears in the loss term due to collisions between
wavepackets of the two different components is $\sigma = 4\pi a_0^2$,
as expected from two-body scattering theory for different spin
components.  In the SVEA derivation this follows from the fact that
the cross phase modulation terms are then of the form
$-\frac{i}{\hbar} U_0 N_{T} (|\Phi^b_2|^2) \Phi^a_1$, rather than
$-\frac{i}{\hbar} U_0 N_{T} (2|\Phi^a_2|^2) \Phi^a_1$ as here (note
the factor of $2$), where the superscripts $a$ and $b$ denote spin
indices.

\begin{figure}[tb]
\centerline{\epsfxsize=3.0in\epsfbox{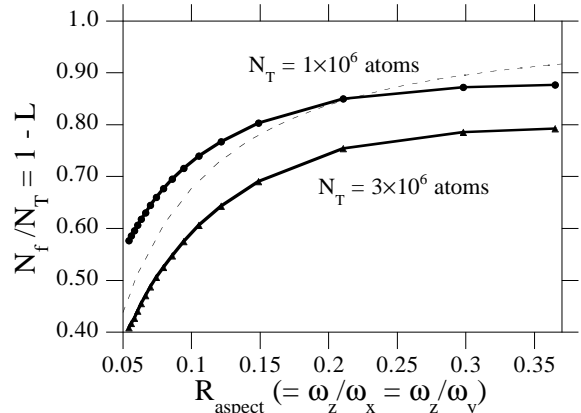}}
\caption {Fraction of atoms remaining in the condensate wavepackets
after the wavepackets have separated.  A fast daughter wavepacket with
half the initial number of atoms and with momentum $2\hbar k$ in the
$z$-direction moves through the remaining parent condensate with zero
central momentum.  The dashed line shows the result of the heuristic 
model for $10^6$ atoms.}
\label{f2}
\end{figure}

As a first example, we consider a condensate of Na atoms in the $F=1$,
$M=-1$ Zeeman sublevel in a cigar-shaped trap elongated in the
$z$-direction.  After the trapping potential is turned off, the
condensate is allowed to freely evolve for 600 $\mu$s, and a short
duration Bragg scattering pulse is applied that creates a $2\hbar{\bf
k}_{ph} = (2h/\lambda_{ph}) {\hat z}$ momentum component.  We consider
the case with half the initial atoms in the high momentum component
and half in the parent condensate.  We evolve the condensate
wavepackets using a full 3D implementation of
Eqs.~(\ref{SVEA1})-(\ref{SVEA2}) until the wavepackets move apart and
are physically separated.  Fig.~(\ref{f2}) shows $N_f/N_T$ versus the
aspect ratio, ${\rm R}_{aspect}$, of the initial elliptically shaped
BEC for two different initial total number of atoms in the BEC, $N_T =
1.0\times 10^6$ atoms and $N_T = 3.0\times 10^6$ atoms respectively. 
Here $N_f$ is the total number of atoms remaining in both the
$0\hbar{\bf k}_{ph}$ and $2\hbar{\bf k}_{ph}$ wavepackets after the
wavepackets separate.  Thus, $N_f/N_T = 1 - L$ where $L$ is the
fractional loss of atoms from the mean-field due to elastic
scattering.  The Thomas-Fermi aspect ratio is related to the trap
frequencies by ${\rm R}_{aspect} \equiv x_{TF}/z_{TF} =
\omega_z/\omega_x$.  The actual trap frequencies in our calculation
were $\nu_z = 30.7$ Hz, and $\nu_x =\nu_y = \nu_z/{\rm R}_{aspect}$. 
The figure shows that the loss increases as the aspect ratio
decreases, and as the total number of atoms increases, reaching 60\%
for 3 million atoms and ${\rm R}_{aspect} \approx 1/20$.  $N_f/N_T$
rises slowly to unity as ${\rm R}_{aspect}$ gets large.

A simple heuristic model helps to explain the magnitude of the losses. 
Assume a uniform atom density of $n=N_T/V$ in a cylinder of length
$\ell=2z_{TF}$, area $A=\pi x_{TF}^2$, and volume $V=\ell A$.  If an
equal number of atoms $N_T/2$ is assumed to be in the daughter and
parent wavepackets, and depletion of $n$ during the interaction is
ignored, a simple argument shows that the fraction of atoms remaining
after the packets separate is
$(\ell_{mfp}/\ell)(1-e^{-\ell/\ell_{mfp}})$, where
$\ell_{mfp}=(\frac{n}{2}\sigma)^{-1}$ is the mean-free-path for the
collision.  For example, with $\nu_z=30.7$ Hz, ${\rm R}_{aspect}=0.1$,
and $N_T=10^6$ atoms, we find that $\ell=120$ $\mu$m and
$\ell_{mfp}=140$ $\mu$m are comparable in magnitude.  The dashed line
in Fig.~\ref{f2} shows that this simple model qualitatively accounts
for our results.

\begin{figure}[tb]
\centerline{\epsfxsize=3.0in\epsfbox{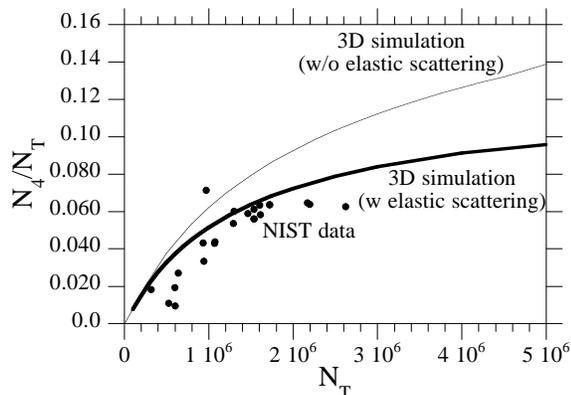}}
\caption {Fraction of atoms in the 4WM output wavepacket, $N_{4}/N_T$,
versus the number of initial atoms, $N_T$.  The dots represent
experimental data {\protect \cite{4WM_2}}, the solid curve is the 3D
calculation without elastic scattering, and the heavy solid curve is
the 3D calculation with elastic scattering.}  
\label{f3}
\end{figure}

In the NIST 4WM experiment \cite{4WM_2}, the Na$(F=1,M=-1)$ condensate
is exposed to Raman scattering pulses which create copies of the
parent condensate at central momenta $\hbar{\bf k}_2 =
(h/\lambda_{ph}) ({\hat x}+{\hat y})$ and $\hbar {\bf k}_3 =
(2h/\lambda_{ph}) {\hat x}$, leaving part of the atoms in the
$\hbar{\bf k}_1 = {\bf 0}$ wavepacket.  The treatment of elastic
scattering from the disparate momentum components of the wavepacket in
the 4WM experiment is similar to the description above for the two
momentum component case.  Now, there are three elastic scattering loss
terms for each SVE momentum component $\Phi_j$ arising due to the
cross-phase modulation terms of each momentum component with the other
three momentum components.  We also included the momentum dependent
correction term in Eq.~(\ref{Tmatrix}) in the coupling constant for
the 4WM source terms on the right hand side of Eq.~(\ref{SVEA}); it
only slightly decreases 4WM at large $N_T$.  In the experiment a trap
with $\nu_x = 84$ Hz, $\nu_y = 59$ Hz and $\nu_z = 42$ Hz contained a
Na BEC without a discernible non-condensed fraction.  The condensate
was exposed to Raman scattering pulses 600 $\mu$s after the magnetic
harmonic potential was turned off.  Fig.~\ref{f3} shows the fraction
of atoms in the 4WM output wavepacket as a function of the initial
total number of atoms $N_T$ as determined (1) experimentally
(circles), (2) by calculation without including elastic scattering
loss for a ratio of atoms in the three initial wavepackets of 7:3:7,
and (3) by calculation including elastic scattering.  The effects of
elastic scattering are pronounced for large values of $N_T$, with the
percent loss due to elastic scattering reaching 44\% for 5 million
atoms.  The discrepancy with experiment is reduced significantly by
including loss due to elastic scattering.  Possible remaining sources
of discrepancy include micromotion of the BEC in the
time-orbiting-trap, laser misalignment and a small finite temperature
component of the BEC.

It is useful to put the elastic scattering discussed here into
perspective.  The mean-field wavefunction for the zero temperature BEC
is a symmetric product of identical orbitals, each orbital being a
coherent superposition of momentum wavepackets.  Elastic scattering
between the various momentum components of a BEC results in atoms
which can not be described by the mean-field since the ``modes'' into
which the atoms are scattered (there are an infinite number of
scattering angles, or modes, to scatter into) are not macroscopically
populated \cite{Moore,Marshall}.  The momentum components in the
spherical shell can not be generated using the dynamics of the GPE or
the SVEA equations since neither contains terms that produce such
momentum components.  However, the SVEA equations do allow collision
losses to be treated.  In contrast to the formation of the spherical
shell of elastically scattered atoms, the fourth wave in 4WM is
explicitly generated by the GPE or the SVEA equations.

The scattering of atoms into the spherical shell and the loss of atoms
from the condensate are a result of Hamiltonian dynamics; no
interactions with a bath, and therefore no incoherent processes
described by $T_1$ or $T_2$ relaxation times, are necessary.  Our
treatment of the process has, for convenience, used an imaginary
potential that serves as a mechanism to take atoms that are
elastically scattered out of the condensate.  It must also be noted we
were able to carry out our procedure for modeling the loss of atoms
from the condensate only as a result of making the slowly varying
envelope approximation, by which we could track the density of atoms
in each momentum wavepacket individually.  As the elastic scattering
loss increases, further scattering of the elastically scattered atoms
with the condensate atoms will become increasingly important.  This
mechanism is not included in our treatment and would require following
the dynamics of the elastically scattered atoms in detail.

Useful discussions with Carl J. Williams, Eite Tiesinga and Mark
Edwards are gratefully acknowledged.  This work was supported in part
by grants from the Office of Naval Research, the US-Israel Binational
Science Foundation, the James Franck Binational German-Israeli Program
in Laser-Matter Interaction, the National Science Foundation through a
grant for the Institute for Theoretical Atomic and Molecular Physics
at Harvard University and Smithsonian Astrophysical Observatory, and
the National Research Council.

\end{multicols}


\begin{references}
\bibitem{mom_considerations}
The momentum components in the spherical shell cannot be formed by
substituting Eq.~(\ref{in_con}) into $|\Psi|^2 \Psi$, or by any 
other terms in Eq.~(\ref{GP}).
\end{references}
\end{document}